\documentclass{PoS}
\usepackage{graphicx,cite}
\usepackage{epsfig,bm,amsmath}

\usepackage{color}

\usepackage[T1]{fontenc} 
\newcommand{\pa}{\mbox{\scriptsize\sf a}}    
\newcommand{\pb}{\mbox{\scriptsize\sf b}}    
\newcommand{\pq}{\mbox{\scriptsize\sf q}}    
\newcommand{\as}{\alpha_s}                   

\newcommand{\kk}{{\bf k}}            
\newcommand{\ku}{{\bf k}_1}
\newcommand{\kd}{{\bf k}_2}

\newcommand{\dif}{{\rm d}}                   
\newcommand{\du}{\dif[\ku]}\newcommand{\dd}{\dif[\kd]}
\newcommand{\dk}{\dif[\kk]}

\newcommand{\G}{{\cal G}}                    
\newcommand{\e}{\varepsilon}                 

\renewcommand{\o}{\omega}

\newcommand{\beq}{\begin{equation}}
\newcommand{\eeq}{\end{equation}}
\newcommand{\bea}{\begin{eqnarray}}
\newcommand{\eea}{\end{eqnarray}}

\title{Heavy quark impact factor for the LHC phenomenology}

\ShortTitle{Heavy quark impact factor}

\author{Grigorios Chachamis \\
Instituto de F\'{\i}sica Corpuscular, Universitat de Val\`encia -- 
Consejo Superior de Investigaciones Cient\'{\i}ficas, 
Parc Cient\'{\i}fic, E-46980 Paterna (Valencia), Spain \\
E-mail: \email{grigorios.chachamis@ific.uv.es}}

\author{\speaker{Michal De\'ak} \\
Instituto de F\'{\i}sica Corpuscular, Universitat de Val\`encia -- 
Consejo Superior de Investigaciones Cient\'{\i}ficas, 
Parc Cient\'{\i}fic, E-46980 Paterna (Valencia), Spain \\
E-mail: \email{michal.deak@ific.uv.es}}

\author{Germ\'an Rodrigo \\
Instituto de F\'{\i}sica Corpuscular, Universitat de Val\`encia -- 
Consejo Superior de Investigaciones Cient\'{\i}ficas, 
Parc Cient\'{\i}fic, E-46980 Paterna (Valencia), Spain \\
E-mail: \email{german.rodrigo@csic.es}}

\abstract{
We comment on the calculation of the finite part of the heavy quark impact 
factor 
at next-to-leading logarithmic (NL$x$) accuracy. The result is presented in a 
form suitable for phenomenological 
studies such as the calculation of the cross-section for single heavy quark 
production at the LHC within the $k_T$-factorization scheme.}

\FullConference{The European Physical Society Conference on High Energy Physics -EPS-HEP2013\\
		18-24 July 2013\\
		Stockholm, Sweden}

\begin{document} 

\maketitle

\flushbottom

\section {Introduction}

Significant developments in the last two decades in 
small-$x$ physics made possible the
phenomenological 
analysis of deep inelastic scattering (DIS) and other high energy scattering processes 
within the $k_T$-factorization scheme. 
They were mainly driven by the Balitsky-Fadin-Kuraev-Lipatov (BFKL) framework
for the resummation of  high center-of-mass energy logarithms
at leading (L$x$)~\cite{BFKLLO} and next-to-leading (NL$x$)~\cite{BFKLNLO}
logarithmic accuracy.

A
key ingredient for studying high energy scattering processes within the
$k_T$-factorisation scheme
is the impact factor, a process dependent object.
The impact factors for gluons and massless quarks have been
calculated in Ref.~\cite{Ciafaloni:1998hu}, at NL$x$
accuracy.
This allows for the calculation of various 
processes with massless quarks and gluons in the initial state. 
The generalization to hadron-hadron collisions has also  
been established~\cite{Mueller:1986ey,Vera:2007kn,Kwiecinski:2001nh}.

The NL$x$ impact factor for a massive quark in the initial state has been calculated
in Ref.~\cite{Ciafaloni:2000sq}. However, the result was
given in the form of a sum of an infinite number of terms. 
To make the result of Ref.~\cite{Ciafaloni:2000sq} available for
phenomenological studies we recalculate the NL$x$ heavy quark 
impact factor in a compact and resummed form which is more suitable 
for numerical applications.

\section{High energy factorisation}

In the high energy limit: $\Lambda_{QCD}\ll |t|\ll s$, the 
partonic cross-section of 
$2\rightarrow 2$ processes factorises into the impact factors $h_{\pa}(\ku)$ and 
$h_{\pb}(\kd)$ of the two colliding partons $\pa$ and $\pb$,
and the gluon Green's function $\G_{\o}(\ku,\kd)$ (here in Mellin space) so
that the differential cross-section can be written as
\begin{equation*}
 \frac{\dif\sigma_{\pa\pb}}{\du\,\dd}=
 \int\frac{\dif\o}{2\pi i\, \o}\, h_{\pa}(\ku) \, \G_{\o}(\ku,\kd) \,
 h_{\pb}(\kd)\left(\frac{s}{s_0(\ku,\kd)}\right)^{\o}~,
\label{fatt}
\end{equation*}
where $\o$ is the dual variable to the rapidity $Y$, and
$\dk=\dif^{2+2\e}\kk/\pi^{1+\e}$ is the transverse space measure.
The L$x$ quark impact factor 
can be expressed by a very simple formula:
\begin{equation*}
 h^{(0)}({\kk})=\sqrt{\frac{\pi}{N_c^2-1}}\;
 \frac{2C_F \as N_{\e}}{\kk^2\,\mu^{2\e}}~, \quad
 N_{\e} = \frac{(4\pi)^{\e/2}}{\Gamma(1-\e)}~,
\label{hzero}
\end{equation*}
and it is the same (up to a color factor) for quarks and gluons. Variables
$\mu$ and $\e$ are the renormalization scale and the dimensional
regularization parameter respectively.

\section{The analytic impact factor}

Collecting all the contributions, the impact factor of a heavy quark 
at NL$x$ accuracy reads 
$h_{\pq}({\bf k}) = h^{(0)}({\bf k}) + h_{\pq}^{(1)}({\bf k})~$, 
and can be expressed in terms of a singular and a finite contribution
\begin{equation*}\label{eq:h(1)}
h_{\pq}({\bf k}) = h_{\pq}^{(1)}({\bf k})|_{\rm sing} + h_{\pq}({\bf k})|_{\rm finite}~.
\end{equation*}
The singular term $h_{\pq}^{(1)}({\bf k})|_{\rm sing}$ is given in~\cite{Ciafaloni:2000sq}.

The finite contribution, which is our main result, 
finally reads 
\begin{equation*}
\begin{split}
h_{\pq}({\bf k})|_{\rm finite} & =
h^{(0)}({\bf k},\alpha_s({\bf k})) \, \Bigg\{1 + \frac{\as \, N_c}{2\pi} \, \Bigg[ 
\mathcal{K}-\frac{\pi^2}{6} + 1 - R \, \log(4R)    
- \log(Z) \left(1+2R\right)\sqrt{\frac{1+R}{R}}\\ & - 2 \log(Z)^2 
-  3 \, \sqrt{R} \left({\rm Li}_2(Z)-{\rm Li}_2(-Z)
+ \log(Z) \, \log \left(\frac{1-Z}{1+Z} \right) \right) \\
& + {\rm Li}_2\left(4R\right) \, \Theta_{m\, k} + \left(
  \frac{1}{2} \log\left(4R\right)+\frac{1}{2} \log^2\left(4R\right)
+ {\rm Li}_2 \left( \frac{1}{4R} \right) \right) \, \Theta_{k\,m}  
\Bigg] \Bigg\}~,
\label{eq:finalfinite}
\end{split}
\end{equation*}
with $R={\bf k}^2/(4m^2)$ and $Z=(\sqrt{1+R}+\sqrt{R})^{-1}$.
As in Ref.~\cite{Ciafaloni:2000sq}, we have absorbed the singularities 
proportional to the beta function into the running 
of the strong coupling $\alpha_s({\bf k})$~\cite{Rodrigo:1993hc}, for  details
we refer the reader to Ref.~\cite{hqif1}.
The heavy quark impact factor, as described here, will be applied to high-energy
phenomenology study of the cross section for
single heavy quark forward production at the LHC~\cite{hqif2}. 

\section*{Acknowledgments}

This work has been supported by the 
Research Executive Agency (REA) of the European Union under 
the Grant Agreement number PITN-GA-2010-264564 (LHCPhenoNet),
by the Spanish Government and EU ERDF funds 
(grants FPA2007-60323, FPA2011-23778 and CSD2007-00042 
Consolider Project CPAN) and by GV (PROMETEUII/2013/007). 
GC acknowledges support from Marie Curie Actions (PIEF-GA-2011-298582).
MD acknowledges support from Juan de la Cierva programme (JCI-2011-11382).


\begin{thebibliography}{10}

\bibitem{BFKLLO}  
L.~N.~Lipatov, 
Sov.\ J.\ Nucl.\ Phys.\  {\bf 23} (1976) 338;
%
E.~A.~Kuraev, L.~N.~Lipatov, V.~S.~Fadin,
Phys.\ Lett.\  B {\bf 60} (1975) 50, 
Sov.\ Phys.\ JETP {\bf 44} (1976) 443,
Sov.\ Phys.\ JETP {\bf 45} (1977) 199;
%
Ia.~Ia.~Balitsky, L.~N.~Lipatov, 
Sov.\ J.\ Nucl.\ Phys.\  {\bf 28} (1978) 822. 


\bibitem{BFKLNLO}
V.~S.~Fadin, L.~N.~Lipatov, Phys.\ Lett.\  B {\bf 429} (1998) 127 [hep-ph/9802290];
%
 M.~Ciafaloni, G.~Camici, Phys.\ Lett.\  B {\bf 430} (1998) 349 [hep-ph/9803389].
  

\bibitem{Ciafaloni:1998hu} 
  M.~Ciafaloni and D.~Colferai,
  Nucl.\ Phys.\ B {\bf 538}, 187 (1999)
  [hep-ph/9806350].
  
\bibitem{Mueller:1986ey}
  A.~H.~Mueller and H.~Navelet,
  Nucl.\ Phys.\ B {\bf 282} (1987) 727.

\bibitem{Vera:2007kn}
  A.~Sabio Vera and F.~Schwennsen,
  Nucl.\ Phys.\ B {\bf 776} (2007) 170
  [hep-ph/0702158 [HEP-PH]].
  
\bibitem{Kwiecinski:2001nh}
  J.~Kwiecinski, A.~D.~Martin, L.~Motyka and J.~Outhwaite,
  Phys.\ Lett.\ B {\bf 514} (2001) 355
  [hep-ph/0105039].

\bibitem{Ciafaloni:2000sq}
  M.~Ciafaloni and G.~Rodrigo,
  JHEP {\bf 0005} (2000) 042
  [hep-ph/0004033].

\bibitem{Rodrigo:1993hc}
  G.~Rodrigo and A.~Santamaria,
  Phys.\ Lett.\ B {\bf 313} (1993) 441
  [hep-ph/9305305];
  G.~Rodrigo, A.~Pich and A.~Santamaria,
  Phys.\ Lett.\ B {\bf 424} (1998) 367
  [hep-ph/9707474].

\bibitem{hqif1}
  G.~Chachamis, M.~Deak and G.~Rodrigo,
  arXiv:1310.6611 [hep-ph];
G.~Chachamis, M.~Deak and G.~Rodrigo,
  arXiv:1307.2780 [hep-ph].
\bibitem{hqif2}
  G.~Chachamis, M.~Deak, M.~Hentschinski, G.~Rodrigo, C.~Salas and  A.~S.~Vera, work in progress.
  
\end{thebibliography}
\end{document}